\documentclass[12pt]{article}
\usepackage[dvips]{graphicx}
\usepackage{amssymb}
\begin{document}
%\twocolumn{}
\begin{center}
{ \LARGE {\bf On The Darboux B\"{a}cklund Transformation of Optical Solitons with Resonant and Nonresonant Nonlinearity
\\\
 } }
\end{center}
\vskip 0pt
\begin{center}
{\it {\large $Arindam\hskip 2 pt Chakraborty$ \\
 {Department of Physics,Heritage Institute of Technology,Kolkata-700107,India} \\
 and \\
 
  $A. Roy \hskip 3 ptChowdhury
$\footnote {(Corresponding author)e-mail: asesh\_r@yahoo.com}  }\\
\it{High Energy Physics Division, Department of Physics ,
                             Jadavpur University,\\
                              Kolkata - 700032,
                                    India}  }
\end{center}

\vskip 20pt
\begin{center}
{\bf Abstract}
\end{center}

\par Solitons in nonlinear optics holds a special role both in theoretical and experimental studies. Several types of evolution equations are seen to govern different situation of physical relevance. One such is the existence of both resonant and nonresonant situation in optical fibre. The corresponding evolution equation was devised by Doktorov et. al., which consists of a forced NLS equation along with two other equations for population difference and polarization. Here, we have followed an earlier formulation of Neugebauer for Darboux-B\"{a}cklund transformation for this coupled systems. This formalism has the advantage that one can write the N-soliton solution altogether.An important difference with the usual non-linear system is that all the field variables are not present in both part of Lax operator. So we are to apply the DT to both part separately.

 \par PACS Number(s):  05.45.Pq, 05.45.Ac, 05.45.-a.

\section{Introduction}
The possibility of existence of soliton in a medium which has both resonant and non-resonant$^1$ (non-inertial Kerr type)nonlinearities$^2$ has opened up a hole new set of nonlinear pde's which are worth studying. It has been shown by Vlasov et. al.$^3$ that soliton in such situation can be generated by the scanning of light beams over the nonlinear medium$^4$ surface provided beam diffraction occurs. It is actually an instance of nonuniform electrodynamic event in nonlinear optics. Suppose in the nonlinear medium $E$, $P$, and $N$ stands respectively for the electric field, polarization and population difference. Also $\eta$ for the refractive index and written as $\eta=\eta_0+t^2\mid E\mid^2$ which includes Kerr effect. By suitable scaling of these three variables, the nonlinear equations for the pulse propagation was found to be
\begin{eqnarray}
e_t+e_z+ie_{xx}+2i\beta^2\mid e \mid^2 e=-\rho\nonumber\\
\rho_t+i\delta\rho=e n\nonumber\\
n_t=-\frac{1}{2}(e\rho^{\ast}+e^{\ast}\rho)
\end{eqnarray}
with
\begin{eqnarray}
E=\frac{\hbar\omega_p}{\mu}e\nonumber\\
P=-i\mu N_0\rho\nonumber\\
N=N_0 n
\end{eqnarray}
It was observed in the reference(3) that under stationary condition at the surface
\begin{eqnarray}
e(t,x,z=0)=e_o(x-vt)\nonumber\\
v_0=\sqrt{\frac{2k_0\eta_0}{\omega_pc}}V_0
\end{eqnarray}
whence equation(1) reduces to
\begin{eqnarray}
e_{z^{\prime}}-v_0e_{x^{\prime}}+ie_{x^{\prime} x^{\prime}}+2i\beta^2\mid e \mid^2 e=\rho\nonumber\\
-v_0\rho_{x^{\prime}}+i\delta\rho=en\nonumber\\
v_0n_{x^{\prime}}=\frac{1}{2}(e\rho^{\ast}+e^{\ast}\rho)
\end{eqnarray}
depending on two variables $(x^{\prime}, z^{\prime})$.
It is now important to observe equation(4)has got a Lax pair written as
\begin{eqnarray}
\Psi_x=U\Psi\nonumber\\
\Psi_z=V\Psi
\end{eqnarray}
with
\begin{equation}
U=\beta e Z_{+}-\beta e^{\ast} Z_{-}-2i\zeta Z_3
\end{equation}
and
\begin{eqnarray}
V=-\beta\left[ie_x+\left(2\zeta-\frac{1}{2\beta}\right)e-\frac{i\rho}{2(\zeta+\beta\delta)}\right]Z_{+}\nonumber\\
-\beta\left[ie^{\ast}_x-\left(2\zeta-\frac{1}{2\beta}\right)e^{\ast}-\frac{i\rho^{\ast}}{2(\zeta+\beta\delta)}\right]Z_{-}\nonumber\\
+i\left[\frac{\beta}{\zeta+\beta\delta}n-2\beta^2\mid e \mid^2+2\zeta\left(2\zeta-\frac{1}{2\beta}\right)\right]Z_3
\end{eqnarray}
where $Z_{\pm}$ and $Z_3$ are the $su(2)$ generators satisfying the commutations
\begin{eqnarray}
[Z_3, Z_{+}]=Z_{+}
\end{eqnarray}
\begin{eqnarray}
[Z_3, Z_{-}]=-Z_{-}
\end{eqnarray}
\begin{eqnarray}
[Z_{+}, Z_{-}]=Z_3
\end{eqnarray}
Introducing the simplest representation of $su(2)$ we can rewrite the equation(5) with $\Psi=(\psi_1,\psi_2)^T$ as
\begin{eqnarray}
\psi_{1x}+i\zeta\psi_1=\beta e \psi_2\\
\psi_{2x}-i\zeta\psi_2=-\beta e^{\ast} \psi_1
\end{eqnarray}
and
\begin{eqnarray}
\psi_{1z}=A\psi_1+B\psi_2\\
\psi_{2z}=C\psi_1-A\psi_2
\end{eqnarray}
with
\begin{eqnarray}
A&=&\frac{i}{2}\left[\frac{\beta}{\zeta+\beta\delta}n-2\beta^2\mid e \mid^2+2\zeta\left(2\zeta-\frac{1}{2\beta}\right)\right]\nonumber\\
B&=&-\beta\left[ie_x+\left(2\zeta-\frac{1}{2\beta}\right)e-\frac{i\rho}{2(\zeta+\beta\delta)}\right]\nonumber\\
C&=&-\beta\left[ie^{\ast}_x-\left(2\zeta-\frac{1}{2\beta}\right)e^{\ast}-\frac{i\rho^{\ast}}{2(\zeta+\beta\delta)}\right]
\end{eqnarray}
\section{Darboux-B\"{a}cklund transformation }
The Darboux-B\"{a}cklund Transformation is an useful tool for thew construction of multi-soliton states from a seed solution$^5$. At the present moment there exist three different approach for its construction. One is the usual matrix transformation of the seed eigenvector of the Lax operator$^6$, the second one invokes such a transformation by the choice of the pole structure(in the complex $\zeta$ plane)  of the new eigenfunction as in the case Riemann-Hilbert problem$^7$. The third one was proposed by Neugebauer et. al.$^8$ while analyzing the exact solutions of general relativity equation with cylindrical symmetry$^9$. It may be mentioned that it was the observation of Belinsky and Zakharov that a Lax pair can be written down for such systems. Here, we proceed to study and construct multi-soliton solution of equation-(4) with the help of method due to Neugebauer et. al.
\par Let, $e_0$, $\rho_0$, $n_0$ be a known seed solution of the set 2 and let $\psi_0$ be the corresponding Lax eigenfunction solution of equation(15, 16). Then as per reference(8) we write the new eigenfunction as$^{10}$
\begin{equation}
\Psi=\Psi_s\Psi_0
\end{equation}
with
\begin{equation}
\Psi_s=\mu(\zeta)P(z,x,\zeta)
\end{equation}
and
\begin{eqnarray}
P=\sum_{j=1}^{N-1}\zeta^j
\left(\begin{array}{cc}
   Q_j & R_j \\
   S_j & T_j
   \end{array} \right)
+\zeta^N
\left(\begin{array}{cc}
   1 & 0 \\
   0 & 1
   \end{array} \right)
\end{eqnarray}
This transformation should satisfy the following conditions;
\par (i)$P(x, z, \zeta)$ is a polynomial in the spectral parameter $\zeta$.
\par (ii)The new Lax operator should also obey the symmetry condition: $\sigma_1 L \sigma_1=L^T$.
\par (iii)$\det M_N=1$.
\par(iv) The zeros $\zeta=\zeta_j$ and $\zeta=\bar{\zeta_j}$, $j=1, 2, \cdots, N$ of $P(\zeta)$ do not depend on $x$, and $t$.
\par (v) Furthermore, at the zeros of $\det P(\zeta)$ we should have
\begin{eqnarray}
\hat{\phi}(\zeta_j)=b_j\hat{\psi}(\zeta_j)
\end{eqnarray}
with $j=1, 2, \cdots, N$ of $P(\zeta)$. Here,$\hat{\phi}(\zeta_j)$ and $\hat{\psi}(\zeta_j)$ are two sets of solutions of the Lax equations.
In order to ensure that $\Psi$ is a new solution we consider
\begin{eqnarray}
U=P_xP^{-1}+PU_0P^{-1}\nonumber\\
V=P_tP^{-1}+P V_0P^{-1}
\end{eqnarray}
\par (vi) And last but not least the analytic structure of $U$ and $V$ in the $\zeta$ plane remain intact. That is the new $U$ and $V$ should be of the form
\begin{eqnarray}
V&=&D(x, z)+\zeta W(x, z)\nonumber\\
U&=&H(x, z)+\zeta K(x, z)+\frac{L(x, z)}{\zeta+\beta\delta}
\end{eqnarray}
Writing the equation in full weight
\begin{eqnarray}
\left[i\zeta
\left(\begin{array}{cc}
   -1 & 0 \\
   0 & 1
   \end{array} \right)
+\beta
\left(\begin{array}{cc}
   0 & e \\
   -e^{\ast} & 0
   \end{array} \right)\right]\left[\sum_{j=1}^{N-1}\zeta^j
\left(\begin{array}{cc}
   Q_j & R_j \\
   S_j & T_j
   \end{array} \right)
+\zeta^N
\left(\begin{array}{cc}
   1 & 0 \\
   0 & 1
   \end{array} \right)\right]\nonumber\\
=\sum_{j=1}^{N-1}\zeta^j
\left(\begin{array}{cc}
   Q_{jx} & R_{jx} \\
   S_{jx} & T_{jx}
   \end{array} \right)\\
   +
\left[\sum_{j=1}^{N-1}\zeta^j
\left(\begin{array}{cc}
   Q_{j} & R_{j} \\
   S_{j} & T_{j}
   \end{array} \right)+\zeta^N
\left(\begin{array}{cc}
   1 & 0 \\
   0 & 1
   \end{array} \right)\right]\left[i\zeta
\left(\begin{array}{cc}
   -1 & 0 \\
   0 & 1
   \end{array} \right)
+\beta
\left(\begin{array}{cc}
   0 & e_0 \\
   -e^{\ast}_0 & 0
   \end{array} \right)\right]
\end{eqnarray}
which leads to
\begin{eqnarray}
e=e_0+\frac{2i}{\beta}R_{N-1}\\
e^{\ast}=e^{\ast}_0+\frac{2i}{\beta}S_{N-1}
\end{eqnarray}
Let us now go back to equation(20), which explicitly written leads to
\begin{eqnarray}
\left[\left(\begin{array}{cc}
   A & B \\
   C & -A
   \end{array} \right)\right]\left[\sum_{j=1}^{N-1}\zeta^j
\left(\begin{array}{cc}
   Q_j & R_j \\
   S_j & T_j
   \end{array} \right)
+\zeta^N
\left(\begin{array}{cc}
   1 & 0 \\
   0 & 1
   \end{array} \right)\right]\nonumber\\
=\sum_{j=1}^{N-1}\zeta^j
\left(\begin{array}{cc}
   Q_{jz} & R_{jz} \\
   S_{jz} & T_{jz}
\end{array} \right)+\left[\sum_{j=1}^{N-1}\zeta^j
\left(\begin{array}{cc}
   Q_j & R_j \\
   S_j & T_j
   \end{array} \right)
+\zeta^N
\left(\begin{array}{cc}
   1 & 0 \\
   0 & 1
   \end{array} \right)\right]\left[\left(\begin{array}{cc}
   A_0 & B_0 \\
   C_0 & -A_0
   \end{array} \right)\right]
\end{eqnarray}
Equating similar power of $\zeta$ element-wise we get;
\begin{eqnarray}
n=n_0+2\beta^2\delta[\frac{2i}{\beta}(e_0S_{N-1}+e^{\ast}_0R_{N-1})-\frac{4}{\beta^2}R_{N-1}S_{N-1}]\nonumber\\
+2\beta[\frac{2i}{\beta}(e_0S_{N-1}+e^{\ast}_0R_{N-1})-\frac{4}{\beta^2}R_{N-1}S_{N-1}]Q_{N-1}\nonumber\\
-2[i(2\beta\delta-\frac{1}{2\beta})(e_0+\frac{2i}{\beta}R_{N-1})-(e_{0x}+\frac{2i}{\beta}(R_{N-1})_x)]S_{N-1}\nonumber\\
-\frac{2i}{\beta}(Q_{N-1})_z-2[e^{\ast}_{0x}+i(2\beta\delta-\frac{1}{2\beta})e^{\ast}_0]R_{N-1}\nonumber\\
-4i[(e_0+\frac{2i}{\beta}R_{N-1})S_{N-2}+e^{\ast}_0 R_{N-2}]
\end{eqnarray}
and
\begin{eqnarray}
\rho=\rho_0+4i\delta (R_{N-1})_x-\frac{2\delta}{\beta}R_{N-1}\nonumber\\
+2[\beta\{2\mid e_0 \mid^2+\frac{2i}{\beta}(e_0S_{N-1}+e^{\ast}_0R_{N-1})\frac{4}{\beta^2}R_{N-1}S_{N-1}\}+\frac{\delta}{\beta}]R_{N-1}\nonumber\\
-2(4\delta-\frac{1}{\beta^2})R_{N-2}-\frac{8}{\beta}R_{N-3}\nonumber\\
+2[e_{0x}+\frac{2i}{\beta}(R_{N-1})_x-i(2\beta\delta-\frac{1}{2\beta})(e_0+\frac{2i}{\beta}R_{N-1})]T_{N-1}\nonumber\\
-2[e_{0x}-i(2\beta\delta)e_0]Q_{N-1}+4ie_0Q_{N-2}-\frac{2i}{\beta}(R_{N-1})_z-4i(e_0+\frac{2i}{\beta}R_{N-1})T_{N-2}
\end{eqnarray}
with a similar expression for $\rho^{*}$ in terms of $\rho_0^{*}$ and other functions.
So the full transformation will be known if the coefficient functions $Q_{N-1}$, $R_{N-1}$, $Q_{N-2}$ and $R_{N-2}$ can be determined.For this let us go back to equation(21) and consider both sides to be transformed by the polynomial matrix $P(x, z, \zeta)$. These leads to the following set of equations:
\begin{eqnarray}
\sum_{j-1}^{N-1}(Q_j+\nu_i R_j)\zeta^j_i=-\zeta^N_i\nonumber\\
\sum_{j-1}^{N-1}(T_j+\nu^{-1}_i S_j)\zeta^j_i=-\zeta^N_i
\end{eqnarray}
which gives us a set of linear equations for $Q_i, S_i, R_i$ and $T_i$ .
In equation(29)
\begin{eqnarray}
\nu_i=\frac{\psi^0_{21}-b_i\psi^0_{22}}{\psi^0_{11}-b_i\psi^0_{12}}
\end{eqnarray}
By Cramer's rule one can write
\begin{eqnarray}
Q_{N-k}=-\frac{\Delta^Q_{N-k}}{\Delta}\nonumber\\
R_{N-k}=-\frac{\Delta^R_{N-k}}{\Delta}\nonumber\\
S_{N-k}=-\frac{\Delta^S_{N-k}}{\Delta}\nonumber\\
T_{N-k}=-\frac{\Delta^T_{N-k}}{\Delta}
\end{eqnarray}
Where
\begin{eqnarray}
\Delta= \left|\begin{array}{cccccccccc}
   1 & \nu_1 &\zeta_1&\nu_1\zeta_1&\zeta_1^2&\ldots&\zeta_1^{N-1}&\zeta_1^{N-1}\nu_1\\
   1 & \nu_2&\zeta_2&\nu_2\zeta_2&\zeta_2^2&\ldots&\zeta_2^{N-1}&\zeta_2^{N-1}\nu_2\\
   \vdots&\vdots&\ddots&\vdots\\
   1&\nu_{2N}&\zeta_{2N}&\nu_{2N}\zeta_{2N}&\zeta_{2N}^2&\ldots&\zeta_{2N}^{N-1}&\zeta_{2N}^{N-1}\nu_{2N}
 \end{array} \right|
\end{eqnarray}
and
\begin{eqnarray}
\Delta^Q_{N-k}&=&\det(1,\nu_i,\zeta_i,\nu_i\zeta_i,\zeta^2_i,\cdots,\zeta_i^{N-k-1},\nu_i\zeta_i^{N-k-1},\zeta_i^N,\zeta_i^{N-k}\nu_i,\cdots,\zeta_i^{N-1},\zeta_i^{N-1}\nu_i )\nonumber\\
\Delta^R_{N-k}&=&\det(1,\nu_i,\zeta_i,\nu_i\zeta_i,\zeta^2_i,\cdots,\zeta_i^{N-k-1},\nu_i\zeta_i^{N-k-1},\zeta_i^{N-k},\zeta_i^{N},\cdots,\zeta_i^{N-1},\zeta_i^{N-1}\nu_i )\nonumber\\
\Delta^T_{N-k}&=&\det(1,\nu^{-1}_i,\zeta_i,\nu^{-1}_i\zeta_i,\zeta^2_i,\cdots,\zeta_i^{N-k-1},\nu^{-1}_i\zeta_i^{N-k-1},\zeta_i^N,\zeta_i^{N-k}\nu^{-1}_i,\cdots,\zeta_i^{N-1},\zeta_i^{N-1}\nu^{-1}_i )\nonumber\\
\Delta^S_{N-k}&=&\det(1,\nu^{-1}_i,\zeta_i,\nu^{-1}_i\zeta_i,\zeta^2_i,\cdots,\zeta_i^{N-k-1},\nu^{-1}_i\zeta_i^{N-k-1},\zeta_i^{N-k},\zeta_i^{N},\cdots,\zeta_i^{N-1},\zeta_i^{N-1}\nu^{-1}_i )
\end{eqnarray}

\section{Conclusion}
In our above analysis we have deduced the formula for N-soliton solution for coupled nonlinear system of NLS equation and two other evolution equations involving population difference and polarization of the medium. These equations describe the pulse propagation when the present nonlinearity in both resonant and non-resonant type. An important difference with usual nonlinear problem is that all the nonlinear fields are not present in both part of the Lax pair. So the Darboux transformation need to be applied to the two part of the Lax pair separately for the derivation of the N-soliton formulae for all the variables $n$, $\rho$, $\rho^{\ast}$, $e$ and $e^{\ast}$.

\section{References:}

\par[1]  E. V. Doktorov and R. A. Vlasov- Optica Acta.(1983)\textbf{30}223.\\

\par [2] G. P. Agarwal-Nonlinear Fibre Optics.(Elsevier)2013.\\

\par [3] R. A. Vlasov and V. R. Nagibov(1980)Dokl. Akad Nauk Fielbrassk SSR,\textbf{24} \\

\par [4] L. Matulic and J. H. Eberly-Phys. Rev. (1972)\textbf{A6}822\\

\par [5] V. B. Mateev and M. A. Salle-Darboux Transformation and Solitons(Springer, Berlin 1991).\\

\par [6]E. Fan-J. Phys. A(Math. Gen.)\textbf{23}(2000)6925\\

\par [7] M. Wheeler- An Introduction to Riemann-Hilbert Problems and their application \\

\par [8]G. Neugebauer, R. Meiner-Phys. Lett. \textbf{100A}(1984)467\\

\par [9]D. Maison-Phys. Rev. Lett.(1978)\textbf{41}521\\

\par[10]V. Belinsky and V. E. Zakharov Sov. Phys.JETP \textbf{48(b)}1978.

\end{document}